\documentclass[fleqn,usenatbib]{mnras}

\usepackage[T1]{fontenc}
\usepackage{times}

\usepackage{graphicx}	
\usepackage{amsmath}	
\usepackage{amssymb}	
\usepackage{bm}

\newcommand{\md}{\mathrm{d}}
\newcommand{\me}{\mathrm{e}}
\newcommand{\mi}{\mathrm{i}}

\def\p{\partial}

\newcommand{\bJ}{\mathbf{J}}
\newcommand{\bk}{\mathbf{k}}
\newcommand{\bOm}{\mathbf{\Omega}}

\newcommand{\nn}{\nonumber}

 \title[Unified kinetic theory for stellar systems]{Noise and waves: a unified kinetic theory for stellar systems}

\author[C. Hamilton \& T. Heinemann]{
  Chris Hamilton$^{1}$\thanks{E-mail: ch783@cam.ac.uk} and Tobias Heinemann$^{2}$\\
$^1$Department of Applied Mathematics and Theoretical Physics, University of
Cambridge, Wilberforce Road, Cambridge CB3 0WA, UK\\
$^2$Niels Bohr International Academy, Niels Bohr Institute, Blegdamsvej 17, 2100 Copenhagen, Denmark}
\pubyear{2020}

\begin{document}
\label{firstpage}
\pagerange{\pageref{firstpage}--\pageref{lastpage}}
\maketitle


\begin{abstract}
The traditional Chandrasekhar picture of the slow relaxation of stellar systems assumes that stars' orbits are only modified by occasional, uncorrelated, two-body flyby encounters with other stars. However, the long-range nature of gravity means that in reality large numbers of stars can behave \textit{collectively}. In stable systems this collective behaviour (i) amplifies the noisy fluctuations in the system's gravitational potential, effectively `dressing' the two-body (star-star) encounters, and (ii) allows the system to support large-scale density waves (a.k.a. normal modes) which decay through resonant wave-star interactions.  If the relaxation of the system is dominated by effect (i) then it is described by the Balescu-Lenard (BL) kinetic theory. Meanwhile if (ii) dominates, one must describe relaxation using quasilinear (QL) theory, though in the stellar-dynamical context the full set of QL equations has never been presented. Moreover, in some systems like open clusters and galactic disks, both (i) and (ii) might be important. Here we present for the first time the equations of a unified kinetic theory of stellar systems in angle-action variables that accounts for both effects (i) and (ii) simultaneously. We derive the equations in a heuristic, physically-motivated fashion and work in the simplest possible regime by accounting only for very weakly damped waves. This unified theory is effectively a superposition of BL and QL theories, both of which are recovered in appropriate limits. The theory is a first step towards a comprehensive description of those stellar systems for which neither the QL or BL theory will suffice. 

\end{abstract}

\section{Introduction}
\label{sec:introduction}

When star clusters and galaxies form they rapidly reach stable quasi-steady states via the process of violent relaxation \citep{LyndenBell1967,Binney2008}. Thereafter, in the absence of strong external perturbations such as mergers, these stellar systems evolve very slowly. More precisely, for a stable $N$-star system the characteristic evolution timescale is the so-called \textit{relaxation time} $t_\mathrm{relax} \sim Nt_\mathrm{cross}$, where $t_\mathrm{cross}$ is a typical orbital period. Since $N\sim 10^6$ in globular clusters and $N\sim 10^{11}$ in typical galaxies, there is usually a very large separation between $t_\mathrm{relax}$ and $t_\mathrm{cross}$, and the evolution of $f$ is referred to as \textit{secular}.

Mathematically, secular evolution is described by a kinetic equation of the form 
\begin{equation} 
\label{eqn:dfdt_Cft}
\frac{\partial f}{\partial t} = C[f, ...],
\end{equation}
where $f(\mathbf{x},\mathbf{v})$ is the mean field distribution function (DF), defined such that $\rho(\mathbf{x}) = \int  \md \mathbf{v}\, f$ is the mean mass density at position $\mathbf{x}$ ($\mathbf{v}$ is the velocity), and the `...' allows for dependence on other variables. The quantity $C$ on the right hand side of \eqref{eqn:dfdt_Cft} is referred to as the \textit{collision operator}.  Of course, actual physical-contact collisions of stars are extremely rare; instead, $C$ is sourced by correlations between small fluctuations $\delta f$ of the exact microscopic DF around its mean value $f$, and small fluctuations $\delta \Phi$ of the exact gravitational potential around its mean $\Phi$ (\citealt{Chavanis2012b}). Physically, the potential fluctuations $\delta \Phi$ (with characteristic timescale $\sim t_\mathrm{cross}$) nudge stellar orbits away from their mean field trajectories. The basic premise of kinetic theory is that while one cannot hope to calculate the effect of stochastic fluctuations on an arbitrary star's orbit, the accumulation of many such weak nudges, when applied to the entire system of stars, produces a deterministic secular evolution of $f$ on the timescale $\sim t_\mathrm{relax}$.

Unfortunately a self-consistent calculation of the collision operator $C$ is difficult. To avoid it, \citet{Chandrasekhar1942,Chandrasekhar1943a,Chandrasekhar1943b} pioneered the theory of two-body relaxation. In this theory, one first considers a test star undergoing a series of uncorrelated, independent two-body encounters with field stars. Except during these occasional encounters, all stars are assumed to follow straight line trajectories. 
One then adds up all the encounters that the test star experiences as if they were statistically independent, and pretends that $C$ may be calculated simply by treating the entire system as an ensemble of such test stars. 
Chandrasekhar's theory has been extremely influential in stellar dynamics. However, it has two major shortcomings. First, it ignores the quasi-periodic nature of stars' mean field orbits in the potential $\Phi$ (replacing them with straight lines).  To remedy this one must describe phase space not using position and velocity coordinates $(\mathbf{x},\mathbf{v})$ but using angle-action variables $(\bm{\theta},\bJ)$. Second, and most crucially for this paper, the long-range nature of gravity means that stars can behave \textit{collectively}, i.e. fluctuations in the gravitational potential $\delta \Phi$ can be sourced by coherent motions of many stars at once. Many authors have emphasised the importance of these collective effects for determining the evolution of stellar systems (e.g. \citealt{Julian1966,Weinberg1994,Weinberg1998,Fouvry2015,Hamilton2018,Tamfal2020}). Collective behaviour is manifestly absent in Chandrasekhar's theory, which deals only with statistically independent two-body encounters --- thus, a more sophisticated theory is required.

Roughly speaking, collective behaviour affects $\delta \Phi$ in two ways. First, it amplifies (or `dresses') the steady-state noise over a continuous range of frequencies, analogous to the phenomenon of Debye sheilding in an electrostatic plasma. Second, it allows the system to support transient wave modes at a discrete set of modal frequencies\footnote{Collective behaviour can also lead to instabilities --- see e.g. \citet{Goodman1988,Binney2008,Rozier2019} and references therein.  In this paper we consider only stable systems, so that all wave modes are damped, but our results also apply to weakly unstable waves \citep{Rogister1968}.}; in an electron plasma these are Langmuir waves, with wavelength much larger than the Debye length. We now consider each of these contributions to $\delta \Phi$ in turn.

\subsection{Noise}

The exact gravitational potential of a stellar system is always noisy simply because the system is comprised of a finite number of stars $N$.  However, collective effects can significantly amplify this intrinsic Poisson noise \citep{Weinberg1998,Lau2019}. The resulting amplified or `dressed' noise gradually nudges stars off their mean field orbits, driving secular evolution \citep{Weinberg1993,Fouvry2015}. Mathematically, this secular evolution is described by Balescu-Lenard (BL) theory\footnote{First derived by \citealt{Balescu1960,Lenard1960} in plasma physics; the stellar-dynamical equivalent was derived by \citealt{Luciani1987,Heyvaerts2010,Chavanis2012b,Fouvry2018}).}, so that $C$ in equation \eqref{eqn:dfdt_Cft} corresponds to the BL collision operator $C_\mathrm{BL}[f]$, which is a functional of $f$ alone. 

Moreover, Rostoker's principle \citep{Rostoker1964a,Rostoker1964b,Gilbert1968,Lerche1971} tells us that BL theory can be thought of as a theory of `dressed' star-star interactions. In Rostoker's picture relaxation is driven exclusively by independent two-body encounters just like in Chandrasekhar's theory.  The only difference is that the `bare' Newtonian interaction potential $\psi = -G/\vert \mathbf{x} - \mathbf{x}' \vert$ is replaced by an effective `dressed' interaction $\psi^\mathrm{d}$. Said differently, Chandrasekhar's theory is a theory of star-star interactions \textit{in vacuo} while the BL picture is one of star-star interactions mediated via the `dielectric medium' of the other $N-2$ stars.  In \citet{Hamilton2020} we showed how to derive $C_\mathrm{BL}$ in a very simple way using Rostoker's principle. If one ignores collective amplification then the BL theory reduces to Landau theory, which is a theory of `bare' star-star interactions, and ultimately to Chandrasekhar's theory \citep{Chavanis2013a,Chavanis2013b}.

\subsection{Waves}

Stellar systems also support transient wave modes (\citealt{Binney2008}, \S 5). Unlike amplified noise, these are spatially smooth density waves that do not have a statistically stationary amplitude --- instead they decay via Landau damping.  They tend to have characteristic spatial scales comparable to the size of the cluster/galaxy itself. They are easily excited by external perturbations such as a massive flyby encounter \citep{Murali1999,Vesperini2000}, or they can be self-excited by the Poisson noise inherent in the system's gravitational potential \citep{Weinberg2001,Heggie2020}. When discussing relaxation it is usually assumed that these waves, even if they are present, decay on a short timescale $ \vert\gamma\vert^{-1} \ll t_\mathrm{relax} \sim N t_\mathrm{cross}$, where $-\gamma$ is the \textit{Landau damping rate} of the wave (e.g. \citealt{Fouvry2018}, Appendix F).  Because of this the impact of waves upon secular relaxation is typically neglected.

However, this assumption is not always valid. Some damped wave modes in star clusters and galaxies can be very long-lasting, decaying on a timescale $\vert \gamma \vert^{-1} \gtrsim 10^2 t_\mathrm{cross}$ \citep{Weinberg1991,Weinberg1994,Sellwood1998,Ideta2002}. Hence in `intermediate-$N$' systems such as open clusters \citep{Danilov2012,Danilov2013} or the $N=10^3$-body simulations of \citet{Lau2019}, the assumption $\vert \gamma \vert^{-1} \ll t_\mathrm{relax}$ can be violated.  Additionally, waves can be important in systems with very large $N$. For instance, the spiral structure in galactic disks ($N\gtrsim 10^{11}$) has long been posited to consist of a superposition of weakly \textit{unstable} wave modes \citep{Sellwood2014,Sellwood2019}. The interaction of these waves with stars is thought to heat the stellar velocity distribution \citep{Carlberg1985,Binney1988,Griv2006}. Thus there exist many systems in which wave-star interactions can impact the secular evolution significantly.  


If wave-star interactions dominate the evolution of $f$ then one must employ a so-called quasilinear (QL) kinetic theory and take $C$ equal to the QL collision operator $C_\mathrm{QL}[f, \{ \mathcal{E}_g \}]$, where $\{\mathcal{E}_g\}$ is the set of wave energies. In that case one must also prescribe an equation that determines the time-variation of each $\mathcal{E}_g$. In plasma physics there exists a huge literature on the QL kinetic theory of wave-particle interactions (e.g. \citealt{Tsytovich1995,Diamond2010,Ichimaru2018} and references therein). In the gravitational context, a particular QL theory was developed for application to galactic disks and Saturn's rings by \cite{Griv1994,Griv2002,Griv2003,Griv2006}, but that work could not apply to arbitrary geometries since it did not employ angle-action variables. Meanwhile, QL theory in angle-action variables was developed in the context of cylindrical/toroidal plasmas by \citet{Kaufman1970,Kaufman1971,Kaufman1972}. Some aspects of angle-action QL theory, such as the diffusion rate of a test particle and the Landau damping rate for weakly damped modes, have been derived in stellar dynamics \citep{LyndenBell1972, Tremaine1984,Binney1988, Tremaine1998,Nelson1999}. Despite this, a fully consistent set of QL equations in angle-action space has never been written down for stellar systems.

\subsection{Plan for this work}

In reality, both noise and waves are always present. Thus star-star and wave-star interactions are always in effect simultaneously. A general kinetic theory of stellar systems must therefore account for both effects at once.  
In plasma theory several authors have proposed some form of unification of the BL and QL theories --- see for instance \cite{Nishikawa1965,Yamada1965,Klimontovich1967,Rogister1968,Rogister1969,Hitchcock1983,Baalrud2008,Baalrud2010}. However this has never been done the general stellar-dynamical context.

In this paper we present for the first time a unified kinetic theory of stellar systems in angle-action variables which includes both BL and QL theory as limiting cases. 
In the spirit of \citet{Hamilton2020} we keep the derivation as short as possible, employing heuristic techniques and emphasising physical insight throughout (a rigorous derivation is postponed to future work). When considering wave-star interactions we work in the simplest possible limit, accounting only for very weakly damped waves (defined precisely in \S\ref{sec:weakly_damped}), ignoring wave-wave coupling, and making the random phase approximation.  In this limit the BL and QL parts of the theory effectively decouple so that the unified theory is a superposition of the two \citep{Yamada1965,Rogister1968}. We already derived the BL part of the theory in \citet{Hamilton2020} from Rostoker's principle; here we derive the QL part using a similar approach. First we `quantise' each wave into many small pieces (quasiparticles; in plasma theory these are called `plasmons'. Then we account for the `emission' and `absorption' of wave quanta by individual stars using Feynman diagrams, adding up all possible interactions as if they were independent.   The approach we take resembles that used by \citet{Kaufman1971} for cylindrical plasmas, but in our case the mean field has arbitrary geometry (provided it is integrable). To shorten the derivation further we quote several general results from the linear response theory work of \citet{Nelson1999}.

In \S\ref{sec:fundamental} we introduce the notation that we will use for the rest of the paper and write down the fundamental equations of the kinetic theory (though we shall not solve them formally). Then we consider separately dressed noise (\S\ref{sec:noise}) and waves (\S\ref{sec:waves}), and derive their respective contributions to the kinetic theory.  In \S\ref{sec:Discussion} we collect the equations of the unified kinetic theory and discuss how one recovers its BL and QL limits. We conclude in \S\ref{sec:conclusion}.


\section{Fundamental equations}
\label{sec:fundamental}

As in \citet{Hamilton2020}, we consider a system of $N \gg 1$ stars with equal mass $m$, and we assume the mean field potential $\Phi$ is integrable.  Then we introduce angle-action coordinates $(\bm{\theta}, \bJ)$ such that a star's mean field specific Hamiltonian $H = \mathbf{v}^2/2 + \Phi(\mathbf{x})$ does not depend on $\bm{\theta}$, i.e. $H=H(\bJ,t)$, and the angle dependence of all quantities is periodic under ${\theta}_i \to {\theta}_i + 2\pi$. A star's mean field equations of motion are then $\md \bm{\theta}/\md t = \partial H/\partial \mathbf{J} \equiv \bOm(\bJ)$ and $\md \mathbf{J}/\md t = \bm{0}$. Furthermore, by Jeans' theorem the mean field DF of stars in phase space is also a function of actions alone, $f = f(\mathbf{J},t)$. We normalise $f$ so that $\int \md \bm{\theta} \md \mathbf{J} f = \int \md \mathbf{x} \md \mathbf{v} f = Nm$, and so Poisson's equation for the mean quantities reads $\nabla^2 \Phi = 4\pi G \int \md \mathbf{v} f$. Finally, $H$ and $f$ evolve on the timescale $ t_\mathrm{relax} \gg t_\mathrm{cross}$, so from now on we will suppress their $t$ arguments, simply writing $H(\bJ)$, $f(\bJ)$. 

The fundamental equations that govern our exact $N$-body system are the Klimontovich equation and the Poisson equation \citep{Chavanis2012b}:
\begin{align}
  \label{eqn:klim}  &\frac{\md}{\md t}(f+\delta f) = 0, \\
    &\nabla^2 (\Phi+\delta \Phi) = 4\pi G\int \md \mathbf{v} \, (f+\delta f).
    \label{eqn:poisson}
\end{align}
By linearising these equations (i.e. assuming $\delta f \ll f, \, \delta \Phi \ll \Phi$) and waiting a few crossing times\footnote{One must wait for noise to be \textit{synchronised} or \textit{thermalised} \citep{Rogister1968}.  For instance, if the initial conditions are drawn randomly from $f$, so that at $t=0$ the only fluctuations are due to bare Poisson noise (\S\ref{sec:noise}), then it will take a finite time for that bare noise to be `dressed' by collective effects \citep{Lau2019}.  In the plasma analogy, it takes a finite time for an initially random collection of electrons and ions to group into a statistically stationary configuration of Debye polarisation clouds.} one can straightforwardly relate the total potential fluctuation $\delta \Phi$ to the `ballistic fluctuation' $\delta \Phi^\mathrm{ball}$, i.e. the fluctuation that results if one takes the initial phase space density fluctuation $\delta f(t=0)$ and assumes that all stars follow their mean field trajectories indefinitely. They are related via a linear integral operator equation which schematically we may write as:
\begin{align}
\label{eqn:VP}
\hat{\epsilon} \cdot \delta \Phi  = \delta \Phi^\mathrm{ball}, \,\,\,\,\,\,\, \mathrm{where} \,\,\,\,\,\,\,\, \hat{\epsilon} \equiv 1 - 2\pi \hat{\psi} \hat{P},
\end{align}
is the \textit{dielectric operator} which is a functional of $f$.  Here $\hat{\psi}$ is simply the Newtonian interaction $\psi = -G/\vert\mathbf{x}-\mathbf{x}'\vert$ written in operator form, while $\hat{P}$ is the \textit{polarisation operator} which accounts for collective amplification.  We have chosen the normalisation and notation for $\hat{P}$ to coincide with that of  \citet{Nelson1999} --- see their equation (30) --- from whom we will quote various general results throughout this paper. 
An explicit expression for $\hat{P}$ in position-frequency space is given in equation (94) of \citet{Nelson1999}, but will not be required here.  If collective effects are switched off then $\hat{P}=0$. If external perturbations were present they would also feature on the right hand side of \eqref{eqn:VP} alongside $\delta \Phi^\mathrm{ball}$, but we will ignore them here for simplicity.


\section{Dressed noise, star-star interactions, and the Balescu-Lenard collision operator}
\label{sec:noise}

In the simplest possible model of a stellar system, all stars are frozen on to mean field orbits dictated by the smooth potential $\Phi$. This is equivalent to setting the polarization $\hat{P}$ in equation \eqref{eqn:VP} to zero, so that $\delta \Phi = \delta \Phi^\mathrm{ball}$. If we  draw stars' initial phase space locations at random from the mean distribution $f$, then the only potential fluctuations will be due to Poisson noise owing to the finite number $N$ of stars in the system: $\delta \Phi^\mathrm{ball} = \delta \Phi^\mathrm{Poisson}$.  To make this more precise let us write $\delta \Phi$ as a Fourier series:
\begin{align}
\delta \Phi (\mathbf{x}(\bm{\theta},\bJ),t) =\sum_{\bk} \delta \Phi_\bk(\bJ,t) e^{i\bk\cdot \bm{\theta}}.
\end{align}
Here $\bk $ (the `wavenumber') is a dimensionless 3-vector  with integer entries, and the sum is over all such 3-vectors. Then for the system we have just described we would find that the steady-state correlator of potential fluctuations is \citep{Lifshitz1981,Chavanis2012b,Fouvry2018}:
\begin{align}
\label{eqn:corr_poiss}
&\langle \delta \Phi_{\bk} (\bJ,t) \delta \Phi^*_{\bk'} (\bJ',t') \rangle_{\mathrm{Poisson}} \nn \\ & \propto \sum_{\bk'} \int \md \bJ' \vert \psi_{\bk\bk'}(\bJ,\bJ')\vert^2 e^{-i\bk' \cdot \bOm' (t-t')}f(\mathbf{J}').
\end{align}
Here we used the shorthand $\bOm' \equiv \bOm(\bJ')$, and $\psi_{\bk\bk'}$ is the Fourier transform of $\psi$. 
Thus the correlator of the potential fluctuations is (i) proportional to the square of the interaction potential ($\sim \vert \psi_{\bk\bk'} \vert^2$), (ii) oscillates at all possible dynamical frequencies available to the system ($\bk'\cdot\bOm'$), and (iii) is weighted by the number of stars capable of emitting power at those frequencies ($\sim \sum_{\bk'}\int \md \bJ' f(\bJ')$).

Next, let us start with the same initial conditions but this time switch on collective effects (a.k.a. `self-gravity'). That is, we allow each star to move not rigidly in the mean field $\Phi$, but self-consistently in the exact gravitational potential $\Phi + \delta \Phi$. Then $\hat{P}$ is nonzero and $\delta \Phi$ is given by the sum of the inhomogeneous solution and homogeneous solution of \eqref{eqn:VP}. These solutions correspond to dressed noise and wave modes respectively. We now discuss the dressed noise; waves will be discussed in \S\ref{sec:waves}.

\subsection{Dressed noise}

Dressed noise is the name we give to the inhomogeneous solution to \eqref{eqn:VP}, namely $\delta \Phi = \hat{\epsilon}^{-1} \cdot \delta \Phi^\mathrm{ball}$.  
This time, if we measure the correlator of the self-consistent potential fluctuations (and ignoring the homogeneous solution of \eqref{eqn:VP} for now) we find it is again given by \eqref{eqn:corr_poiss} except with the replacement
\begin{align}
\label{eqn:replacement}
\vert \psi_{\bk\bk'}(\bJ,\bJ')\vert^2 \to \vert \psi^\mathrm{d}_{\bk\bk'}(\bJ,\bJ',\bk'\cdot\bOm')\vert^2,
\end{align}
where $\psi^\mathrm{d}_{\bk\bk'}(\bJ,\bJ',\omega)$ is the Fourier-Laplace transform of the dressed interaction potential $\psi^\mathrm{d}$, which in \eqref{eqn:replacement} has been evaluated at the dynamical frequency $\omega = \bk'\cdot\bOm'$. We will not write an explicit expression for $\psi^\mathrm{d}$ explicitly here, but note that it is a conceptually simple exercise in linear response theory \citep{Binney2008,Fouvry2018}. It is non-zero at all dynamical frequencies $\bk' \cdot \bOm'$ accessible to the system, and can be very large at frequencies where $\hat{\epsilon}^{-1}$ is small. With the replacement \eqref{eqn:replacement} we say that the Poisson noise has been \textit{amplified} or \textit{dressed} by collective effects. Since the correlator of potential fluctuations is stationary on timescales shorter than $t_\mathrm{relax}$ (i.e. $\langle \delta \Phi(t)\delta\Phi^*(t')\rangle$ depends only on the time difference $t-t'$) we refer to it as representing \textit{steady-state dressed noise}. In rigorous BL deviations this is precisely the noise spectrum used to compute the collision operator $C_\mathrm{BL}$ \citep{Chavanis2012b,Fouvry2018}.


\begin{figure*}
  \centering
 \includegraphics[width=0.85\linewidth]{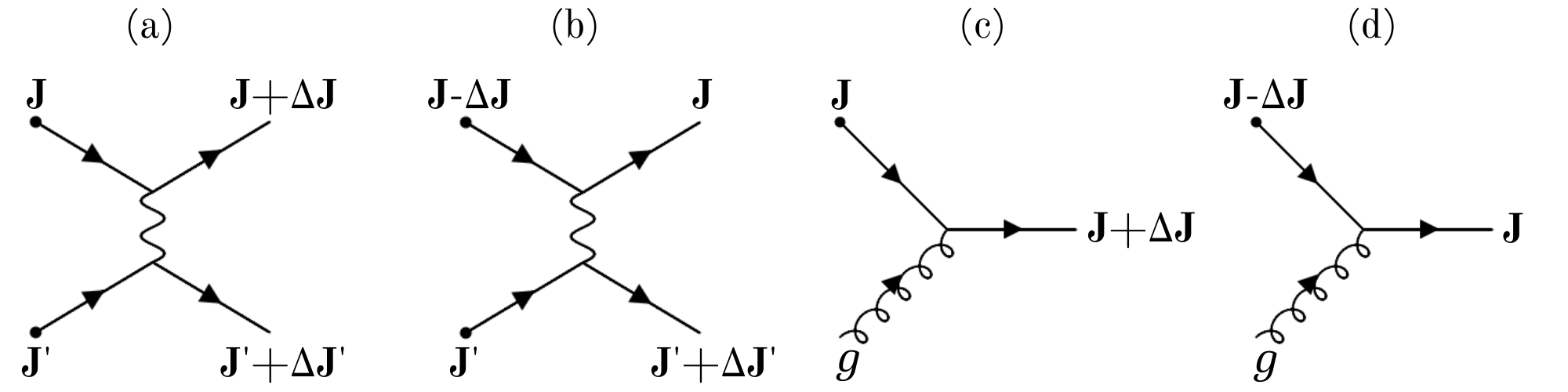}
\caption{Four possibilities for `scattering' in action coordinates. Diagram (a) shows the interaction of a test star with action $\bJ$ and a field star with action $\bJ'$. As a result the test star is kicked to some new action $\bJ + \Delta \bJ$, decrementing the DF $f(\bJ)$.  Similarly, panel (b) shows the test star being kicked \textit{in} to the state $\bJ$, incrementing $f(\bJ)$.  Adding up these processes and their inverses leads to the BL collision operator $C_\mathrm{BL}$ (\S\ref{sec:starstar}).  Analogously, diagrams (c) and (d) show wave-star interactions, whereby a test star is kicked out of (diagram (c)) or in to (diagram (d)) the state $\bJ$ by absorbing a wave quantum (curly line) with pattern frequency $\Omega_g$.  Properly weighted, these processes and their inverses add up to give the QL collision operator $C_\mathrm{QL}$ (\S\ref{sec:wavestar}).  The proper weighting of processes (c) and (d) is summarised by the level diagram in Figure \ref{fig:Level_Diagram}.}
\label{fig:Feynman}
\end{figure*}
\begin{figure}
\centering
\includegraphics[width=0.8\linewidth,trim={0.3cm 0.5cm 0.8cm 0.7cm},clip]{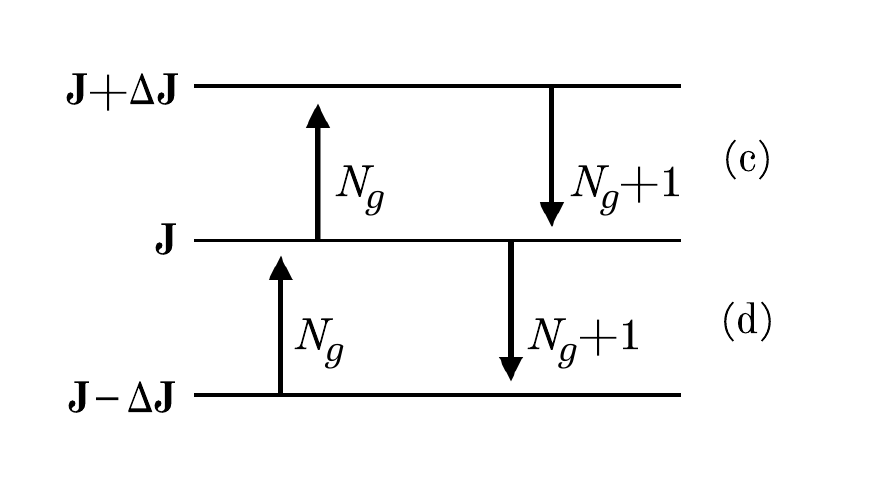}
\caption{Diagram showing the processes of absorption and (stimulated and spontaneous) emission arising from wave-star interactions, as in Figures \ref{fig:Feynman}c,d. A test star's action increases (decreases) by $\Delta \mathbf{J}$ when it absorbs (emits) a wave quantum with frequency $\Omega_g$. While absorption is always stimulated, and hence occurs at a rate proportional to the number of wave quanta $N_g$, emission can be either stimulated or spontaneous and so must be weighted by $N_g+1$ (see \citealt{Thorne2017}, \S 23.3.3).}
\label{fig:Level_Diagram}
\end{figure}


\subsection{Star-star interactions}
\label{sec:starstar}

The interaction of stars with dressed noise drives evolution of $f$. Rostoker's principle tells us that the resulting evolution is mathematically equivalent to that of a superposition of independent two-body (`star-star') encounters, each with effective interaction potential $\psi^\mathrm{d}$. In \citet{Hamilton2020} we showed how to use Rostoker's picture to derive the BL collision operator $C_\mathrm{BL}$. Here we briefly recap that argument.

First we consider a test star with initial phase space coordinates $(\bm{\theta}_0,\bJ)$ interacting with a field star with initial coordinates $(\bm{\theta}_0',\bJ')$.  The Hamiltonian that describes this two-body interaction is 
\begin{align}
\label{eqn:two_body_Hamiltonian}
    h = H(\bJ) + H(\bJ') + m \sum_{\bk \bk'} \me^{\mi(\bk \cdot \bm{\theta} - \bk' \cdot \bm{\theta}') } {\psi}^\mathrm{d}_{\bk \bk'}(\bJ, \bJ', \bk' \cdot \bOm'),
\end{align}
from which we can deduce the first order change in action after time $\tau$, namely $\delta \bJ(\bm{\theta}_0,\bJ,\bm{\theta}_0',\bJ',\tau)$ --- see \citet{Hamilton2020}, equation (3). 

Second we consider the entire stellar system to be an ensemble of test stars, and write down a master equation that accounts for the processes shown in Figure \ref{fig:Feynman}a,b. In words, the distribution function $f(\bJ)$ is decremented if the test star is kicked out of state $\bJ$  by a field star with action $\bJ'$ (Figure \ref{fig:Feynman}a), and incremented if it is instead kicked into $\bJ$ from some other state $\bJ-\Delta \bJ$ (Figure \ref{fig:Feynman}b).  We must also account for the inverses of both diagrams. In each case the test star gets a kick $\Delta \bJ$ and the field star gets a kick $\Delta \bJ'$. We add up all possible interactions by integrating over all field star actions $\bJ'$ and all possible kicks $\Delta \bJ$, $\Delta \bJ'$. When we expand the resulting master equation by assuming that the kicks are small, the terms that survive involve the expectation values $\tau^{-1}\langle \Delta \bJ \Delta \bJ\rangle_\tau$ and $\tau^{-1}\langle \Delta \bJ \Delta \bJ'\rangle_\tau$.

Third we put these pieces together: we use the known first order change $\delta \bJ(\tau)$ and integrate over random phases $\bm{\theta}_0$, $\bm{\theta}_0'$ to calculate $\langle \Delta \bJ \Delta \bJ \rangle_\tau$, and similarly for $\langle \Delta \bJ \Delta \bJ' \rangle_\tau$.  Finally we take the limit $\tau \to \infty$ (since we are interested in the dynamics on a timescale $\gg t_\mathrm{cross}$).  The result is the BL collision operator which is given in \citet{Hamilton2020}, equation (11):
\begin{align}
C_{\mathrm{BL}}[f]
&= \pi (2\pi)^3 m \frac{\p}{\p \bJ} \cdot  \sum_{\bk \bk'} \bk \int \md \bJ' \delta (\bk \cdot \bOm - \bk' \cdot \bOm') 
\nn \\ &\times
\vert \psi^\mathrm{d}_{\bk \bk'}(\bJ, \bJ', \bk' \cdot \bOm') \vert^2
\left( \bk \cdot \frac{\p}{\p \bJ}
-
\bk' \cdot \frac{\p}{\p \bJ'}
\right) f(\bJ) f(\bJ').
\label{eqn:BL}
\end{align}
We see that non-zero contributions to $C_{\mathrm{BL}}$ arise when there is a resonance $\mathbf{k}\cdot \bOm = \bk' \cdot \bOm'$ between the dynamical frequencies of test ($\bk \cdot \bOm$) and field ($\bk' \cdot \bOm'$) stars. The integral over all $\mathbf{J}'$ and sums over $\bk$ and $\bk'$ mean that we scan the entire continuum of dynamical frequencies looking for these resonances. 
Note also that $\psi^\mathrm{d}$ gets evaluated at the dynamical frequency $\mathbf{k}\cdot \bOm$. The amplification due to collective effects will typically be large ($\vert \psi_{\bk\bk'}^\mathrm{d}\vert^2 \gg \vert \psi_{\bk\bk'} \vert^2$) if $\bk\cdot\bOm$ is close to a \textit{modal frequency} $\omega_g$, which we discuss next. 

\section{Wave modes, wave-star interactions, and the quasilinear collision operator}
\label{sec:waves}

Dressed noise is not the only contribution to $\delta \Phi$. The fact that stars behave collectively also allows the system to support \textit{waves} or \textit{normal modes}. These are periodic density oscillations analogous to the fluid modes in a star or planet.  When stellar orbits resonate with a wave they give energy to the wave or absorb energy from it; in a stable system the absorption dominates so the wave \textit{Landau damps} (\citealt{Binney2008}).

Mathematically, waves correspond to homogeneous solutions to the linear operator equation \eqref{eqn:VP}, i.e. they have potential $\delta \Phi_g$ such that $\hat{\epsilon} \cdot \delta \Phi_g = 0$. Thus, the potential of a given wave is an eigenfunction of $\hat{\epsilon}$ corresponding to zero eigenvalue. These eigenfunctions exist for a discrete set of frequencies $\omega_g = \Omega_g + i\gamma_g$, where $g=1,2,...$ is an integer that labels the wave.  Here $\Omega_g \in \mathbb{R}$ is called the \textit{pattern frequency} (not to be confused with the vector of dynamical frequencies $\bOm$), and $-\gamma_g \in \mathbb{R}$ is the Landau damping rate. 
Now, for every wave solution at frequency $\Omega_g + i\gamma_g$ there is another at $-\Omega_g + i\gamma_g$, corresponding to a wave with the same damping rate but travelling in the opposite `direction'.
 For the full homogeneous solution to \eqref{eqn:VP}, which we call $\delta \Phi^\mathrm{w}(\mathbf{x},t)$, to be real the amplitudes of these two waves must be complex conjugates of each other.  With this information it is easy to see that the general homogeneous solution is:
\begin{align}
\delta \Phi^\mathrm{w}(\mathbf{x},t) 
= \sum_g \delta \Phi_g(\mathbf{x},t),
\label{eqn:deltaPhi_waves}
\end{align}
where (c.f. equation (47) of \citealt{Nelson1999})
\begin{align}
\delta \Phi_g(\mathbf{x},t) \equiv A_g(t) \mathrm{Re}[e^{-i\alpha_g}\phi_g(\mathbf{x})  e^{-i\Omega_gt}],
\label{eqn:deltaphiw_NTform}
\end{align}
and crucially the sum in \eqref{eqn:deltaPhi_waves} is limited to waves $g$ such that $\Omega_g > 0$ \citep{Kaufman1971}. The $\phi_g$ in \eqref{eqn:deltaphiw_NTform} are complex spatial `eigenmodes' with units of potential (i.e. energy per unit mass); they are not orthogonal in general. The factor $A_g(t)$ is a real number which is proportional to $e^{\gamma_gt}$, but will have other overall slow time dependence as well. Finally, $\alpha_g$ is an arbitrary phase. 

Since the operator $\hat{\epsilon}$ (and hence its inverse $\hat{\epsilon}^{-1}$) is calculated purely from $f$, waves are a `collisionless' phenomenon in the sense that $\omega_g$ and $\phi_g(\mathbf{x})$ are independent of $N$, and would be the same even in a system with $N\to \infty$. The calculation of $\{ \omega_g, \phi_g(\mathbf{x})\}$ is not the subject of this paper but is a standard, if often difficult, numerical task \citep{Weinberg1994,Binney2008,Heggie2020}.


\subsection{Weakly damped modes}
\label{sec:weakly_damped}

In this paper we will limit our consideration to \textit{weakly damped modes}, by which we mean that the pattern frequency of the wave is much larger than the Landau damping rate:
\begin{align}
\label{eqn:weak_def}
\vert \Omega_g \vert \gg  \vert \gamma_g \vert.
\end{align}
With this assumption, the factor $A_g(t)$ in \eqref{eqn:deltaphiw_NTform} is just a slowly varying real amplitude that includes a factor $e^{\gamma_g t}$, while the fast time dependence of any wave $\sim e^{-i\Omega_g t}$.  

We now write down two standard results that apply to a weakly damped wave with potential $\delta \Phi_g$ (equation \eqref{eqn:deltaphiw_NTform}). First is the wave energy, which contains both the gravitational potential energy of the wave and the kinetic energy of oscillation of the stars that comprise it. Averaged over the period $2\pi/ \Omega_g$, the wave energy is proportional to the square of the wave amplitude (\citealt{Nelson1999}, equation (62)):
\begin{align}
\mathcal{E}_g  = A_g(t)^2 \times \left[- \frac{\pi}{2}\Omega_g \Lambda_g\right],
\label{eqn:Eg}
\end{align}
where
\begin{align}
   \Lambda_g \equiv \int \md \mathbf{x} \, \md \mathbf{x}' \phi^*_g(\mathbf{x}) \phi_g(\mathbf{x}') \frac{\partial P_\mathrm{H}(\mathbf{x},\mathbf{x}',\omega)}{\partial \omega}\bigg\vert_{\omega=\Omega_g},
   \label{eqn:Lambda_g}
\end{align}
is a real normalisation constant with units of $\hbar$ (energy $\times$ time) and can be positive or negative. Here $P_\mathrm{H}$ is the Hermitian part of the polarisation operator $\hat{P}$ expressed in position-frequency space (an explicit expression is given in \citealt{Nelson1999}, equation (96)).
Second is the expression for the Landau growth rate (\citealt{Nelson1999}, equations (63), (64) and (98)):
\begin{align}
  \gamma_g &=  - \frac{(2\pi)^3}{2\Lambda_g}
  \sum_{\bk} \int \md \bJ \,   \vert \phi_{g\bk}(\mathbf{J}) \vert^2  \delta (\bk\cdot \bm{\Omega} - \Omega_g)
 \bk \cdot \frac{\partial f}{\partial \bJ},
 \label{eqn:Landau}
\end{align}
where we have expanded $\phi_g(\mathbf{x}) = \sum_\bk \phi_{g\bk}(\bJ)e^{i\bk\cdot \bm{\theta}}$.
As expected, Landau damping arises when a wave's pattern frequency resonates with the dynamical frequency of a star. 
Note that $\gamma_g$ is independent of the wave amplitude: it does not involve $A_g$, nor does it depend on the normalisation of $\phi_g$ (because of the factor $\Lambda_g^{-1}$).

\subsection{Wave-star interactions}
\label{sec:wavestar}

Waves interact with stars\footnote{They also interact with other waves (`mode coupling').  This nonlinear processes lies beyond the domain of QL theory.}, and these interactions modify both the stellar distribution $f$ and the wave energy. The resulting evolution is the domain of quasilinear (QL) theory. Our primary aim in the rest of this section is to derive the QL collision operator $C_\mathrm{QL}$. 

To proceed we must address a subtlety behind the wave solution \eqref{eqn:deltaPhi_waves}-\eqref{eqn:deltaphiw_NTform}.  The issue is that \eqref{eqn:deltaPhi_waves}-\eqref{eqn:deltaphiw_NTform} is derived by solving \eqref{eqn:VP} as an initial value problem starting at $t=0$. Suppose that at $t=0$ the intial conditions are drawn randomly from $f$, so the only fluctuations in the gravitational potential are from discreteness noise. Then equations \eqref{eqn:deltaPhi_waves}-\eqref{eqn:deltaphiw_NTform} tells us that immediately after $t=0$ there are waves present in the system. Physically these waves are seeded by the specific initial noise fluctuation. By analogy with the plasma literature, we say that the stellar system `emits' waves by virtue of its inherent discreteness.

However, if we were to take \eqref{eqn:deltaPhi_waves}-\eqref{eqn:deltaphiw_NTform} at face value we would conclude that waves with frequency $\omega_g$ will have decayed to a negligible fraction of their initial amplitude by, say, the time $t = 3\gamma_g^{-1}$. After this time there would be effectively no more wave-star interactions to drive the evolution of $f$. However, this would be to ignore the fact that the specific discreteness noise in the system is always fluctuating (though it is of course stationary on average).  Because of this the system continuously emits new waves at frequency $\omega_g$ which are not in general in phase with each other. So while the initial wave may have `died' after $t = 3\gamma_g^{-1}$, in the mean time many other waves of the same frequency will have been born.
 Said differently, there ought to be nothing special about choosing the initial time to be $t=0$ rather than, say, $t=2t_\mathrm{cross}$, or $t=7.34t_\mathrm{cross}$, or whatever, as long as those times are much smaller $\vert \gamma_g \vert^{-1}$, the timescale on which the wave distribution changes significantly. For a more extensive discussion of this point, see \S3 of \citet{Nishikawa1965}.

 Thus, the wave ensemble at each frequency $\omega_g$ really consists of many waves with different amplitudes and phases. The problem is that we do not know these amplitudes or phases individually. However, we can get round this problem by adopting a standard trick from plasma kinetics \citep{Pines1962,Wyld1962,Kaufman1971, Tsytovich1995}. We replace the complicated set of waves at frequency $\omega_g$ with a large number $N_g \gg 1$ of discrete \textit{wave quanta} (known in plasma theory as \textit{quasiparticles} or \textit{plasmons}) of equal amplitude but randomly distributed phase.  Precisely, we label the wave quanta $j=0,1,2,.., N_g$ and assign them amplitudes $A_g^j(t) \equiv A_g(t)/\sqrt{N_g}$ and phases $\alpha_g^j \equiv 2\pi j/N_g$.   Obviously, the total energy in these waves then works out to be $- A_g(t)^2 \pi \Omega_g \Lambda_g/2 = \mathcal{E}_g(t)$, equivalent to that of a single wave with amplitude $A_g(t)$ --- see equation \eqref{eqn:Eg}.  
 
 As we will see, the total wave energy $\mathcal{E}_g$ is a well-defined quantity that obeys a conservation law (\S\ref{sec:energy}).  We will also see that the diffusion rate of stars by wave-star interactions is proportional to $\mathcal{E}_g$, so it is important that we can calculate it.  We choose the integer $N_g$ such that\footnote{Note that we are still pursuing a purely classical calculation. The introduction of Planck's constant $\hbar$ is merely to discretise the wave energy into very small pieces so that $N_g \gg 1$ can be well-approximated by an integer; any small constant with the dimensions of $\hbar$ will do.} $\mathcal{E}_g = N_g\hbar \Omega_g \sigma_g$, where $\sigma_g = -\mathrm{sgn}(\Lambda_g) = \pm 1$ is chosen so that $N_g$ is positive. Then to determine $\mathcal{E}_g(t)$ we simply require knowledge of $N_g(t)$.  We will derive an equation for $N_g(t)$ in \S\ref{sec:wavekinetic}; for now we simply note that from \eqref{eqn:Eg} we have:
\begin{align}
    N_g = -\frac{\pi}{2}\frac{A_g^2 \Lambda_g}{\hbar}\sigma_g = \frac{\pi}{2}\frac{A_g^2 \vert \Lambda_g \vert}{\hbar} \propto e^{2\gamma_g t}.
    \label{eqn:Ng_def}
    \end{align}

Having introduced this quantised picture, we now consider the interaction of stars with wave quanta. Since the number of wave quanta $N_g$ is not conserved, the simplest wave-star interactions possible are the `absorption' or `emission' of a wave quantum by a star as illustrated in Figure \ref{fig:Feynman}c,d. All other interactions are non-linear and so do not enter our theory \citep{Tsytovich1995}.   To write down the master equation for $C_\mathrm{QL}$ we sum these processes and their inverses over all wave labels $g$.  Crucially, we must take care to weight the sums properly. The processes shown in (c) and (d) both correspond to the absorption of a wave quantum by a star. On the other hand, the \textit{inverse} of these processes --- namely emission of a wave by a star ---  can be either stimulated (owing to the fact that the star sits in a bath of waves) or spontaneous (arising just from the discreteness noise produced by the star).  While any given microscopic process has the same probability as its inverse, stimulated processes must be weighted by the number of wave quanta $N_g$, while spontaneous processes are independent of $N_g$ and so carry no such weighting.  The weighting of each processes is summarised in Figure \ref{fig:Level_Diagram}. 

We are now ready to write down the master equation. We introduce $ w^g_{\Delta \bJ}(\bJ)$ as the \textit{transition rate density}, defined such that $ w^g_{\Delta \bJ}(\bJ)\, \md \Delta \bJ \,\tau$ is the probability that a given test star with action $\bJ$ is scattered to the volume of phase space $\md \Delta \bJ$ around $\bJ + \Delta \bJ$ by a given wave quantum with frequency $\Omega_g^{-1}$ after a time interval $\tau$, where $t_\mathrm{cross}, \Omega_g \ll \tau \ll \vert \gamma_g\vert^{-1}$.  Then the master equation for $C_\mathrm{QL}$ reads:
\begin{align}
&C_\mathrm{QL} = 
  \int  \md \Delta \bJ \nn \\
&\times \sum_g   \Big\{  w^g_{\Delta \bJ}(\bJ) \big[
 -N_g f(\bJ)  \nn
+ (N_g+1) f(\bJ+\Delta \bJ)
 \big] \nn
 \\
&+
 w^g_{\Delta \bJ}( \bJ - \Delta \bJ) \big[
N_g  f(\bJ-\Delta \bJ)  - (N_g+1)f(\bJ)
 \big]
\Big \} .
\label{eqn:CQL_master}
\end{align}
 (We again emphasise that the sum over $g$ is restricted only to waves with $\Omega_g > 0$). In each factor $N_g+1$ in this equation, the `$N_g$' term corresponds to stimulated emission while the `$1$' term corresponds to spontaneous emission. The other terms represent absorption.  Next we assume that $\Delta \bJ$ is small compared to $\bJ$ and thereby expand the right hand side. Throwing away terms higher than second order and using $N_g \gg 1$ we get
\begin{align}
\label{eqn:CQL_drag_diffusion}
 C_\mathrm{QL}
=
\frac{\partial}{\partial \bJ}\cdot 
\sum_g 
\left[ 
\mathbf{F}_g f(\bJ) 
+ N_g \mathsf{D}_g \cdot \frac{\partial f}{\partial \bJ} \right], 
\end{align}
where 
\begin{align}
\label{eqn:average_def}
 \mathbf{F}_g
 &=
 \int \md \Delta \bJ \,  w^g_{\Delta \bJ}(\bJ) \,\Delta \bJ \equiv \frac{ \langle \Delta \bJ \rangle_\tau^g }{\tau},
 \\
  \mathsf{D}_g
 &=
 \int \md \Delta \bJ \,  w^g_{\Delta \bJ}(\bJ) \,\Delta \bJ \Delta \bJ \equiv \frac{ \langle \Delta \bJ \Delta\bJ \rangle_\tau^g }{\tau},
\end{align}
and
$\langle Q \rangle^g_\tau$ is the average increment in the quantity $Q$ after time $\tau$ (c.f. equation (6) of \citealt{Hamilton2020}).
Note that the $3$-vector $ \mathbf{F}_g$ is independent of $N_g$; this is a dynamical friction term representing the `recoil' that stars feel when they spontaneously emit wave quanta. Meanwhile the $3\times 3$ matrix $\mathsf{D}_g$ comes multiplied by $N_g$. This term represents the diffusion of stars induced by the bath of random wave quanta. To close the system we must also have an equation that determines $N_g(t)$. This is the so-called \textit{wave kinetic equation}.

\subsection{The wave kinetic equation}

\label{sec:wavekinetic}

To derive the wave kinetic equation we consider the process shown in\footnote{We do not need to consider diagram (d) because we no longer care about keeping track of the star's actions (we are about to integrate over all $\bJ$).} Figure \ref{fig:Feynman}c (wave absorption) and its inverse (wave emission).  We note that the average number of stars in the action space volume element surrounding $\bJ$ is $(2\pi)^3 f(\bJ) \md \bJ /m$.  Then by integrating over all possible stellar actions $\bJ$ and all possible kicks $\Delta \bJ$ we can easily write down a master equation for\footnote{We have implicitly chosen the integer $N_g$ large enough (or equivalently, taken $\hbar$ small enough) that it may be approximated as a continuous variable.} $N_g$:
\begin{align}
&\frac{\partial N_g}{\partial t} \nn \\ &= \frac{(2\pi)^3}{m}\int \md\bJ \, \md \Delta \bJ  \, w^g_{\Delta \bJ}(\bJ)  \big[ - N_g f(\bJ) + (N_g+1) f(\bJ+\Delta \bJ) \big] \nn 
\\
&= \frac{(2\pi)^3N_g}{m} \int \md \bJ \,\mathbf{F}_g \cdot \frac{\partial f}{\partial \bJ} + \frac{(2\pi)^3}{m} \int \md \bJ \,\md \Delta \bJ  w^g_{\Delta \bJ} (\bJ) f(\bJ),
\label{eqn:dNgdt_twoparts}
\end{align}
where to get the second equality we expanded the right hand side for small $\Delta \bJ$ to first order, and used $N_g \gg 1$ and equation \eqref{eqn:average_def}.  

Now, we know from equation \eqref{eqn:Ng_def} that $N_g  \propto e^{2\gamma_gt}$; thus we can identify the first term on the right hand side of \eqref{eqn:dNgdt_twoparts} as $2\gamma_g N_g$, and use the fact that we already have an explicit expression for $\gamma_g$, namely equation\footnote{In \S\ref{sec:cql} we will demand that \eqref{eqn:Landau} be consistent with the first term on the right hand side of \eqref{eqn:dNgdt_twoparts}, thereby fixing $\mathbf{F}_g$.} \eqref{eqn:Landau}. This term represents the loss of wave energy through Landau damping. The second term on the right hand side of \eqref{eqn:dNgdt_twoparts}, which is independent of the wave amplitude, must therefore correspond to the rate of spontaneous emission of wave energy. Hence by multiplying \eqref{eqn:dNgdt_twoparts_2} by $\hbar \Omega_g\sigma_g$ we arrive at the following equation for the wave energy:
\begin{align}
\frac{\partial \mathcal{E}_g}{\partial t} = 2\gamma_g \mathcal{E}_g + S_g,
\label{eqn:dNgdt_twoparts_2}
\end{align}
where $S_g \equiv [(2\pi)^3/m] \hbar\Omega_g \sigma_g \int \md \bJ \md \Delta \bJ \,  w^g_{\Delta \bJ}(\bJ) f(\bJ)$ is the spontaneous emission rate. There are multiple ways to derive an explicit expression for $S_g$. For instance one can calculate the rate at which a single star on its mean field orbit, treated as an external perturber, does work on the system at frequency $\Omega_g$, then sum the contributions from all stars \citep{Kaufman1970,Kaufman1971}.  Alternatively one can demand that the total energy of the system $\mathcal{E}_\mathrm{tot}$ is conserved, and deduce the form of $S_g$ that way.  Here we simply write down the answer:
\begin{align}
S_g &= -\frac{(2\pi)^3m \Omega_g}{\Lambda_g}
\sum_\bk \int \md \bJ \delta(\bk\cdot\bOm - \Omega_g) \vert \phi_{g\bk}(\bJ) \vert^2 f(\bJ).
\end{align}
The fact that the work done on the system goes like $\sim m
\delta(\bk\cdot\bOm - \Omega_g) \vert \phi_{g\bk}\vert^2 f$ might have been guessed a priori. To confirm that the various prefactors are also correct, upon gathering the results of the unified theory we will show that the system's total energy $\mathcal{E}_\mathrm{tot}$ is indeed conserved (\S\ref{sec:energy}).

Putting these pieces together we can now write down the final form of the wave kinetic equation:
 \begin{align}
  \label{eqn:wave_kinetic}
  \frac{\partial \mathcal{E}_g}{\partial t}
  &= -(2\pi)^3 \Omega_g  \sum_\bk \int \md \bJ \, \delta (\bk\cdot \bm{\Omega} - \Omega_g)\frac{\vert \phi_{g\bk}(\mathbf{J}) \vert^2}{\Lambda_g}
  \nn  \\
 &\times
  \left[m f(\mathbf{J}) + \frac{\mathcal{E}_g}{\Omega_g} \, \bk \cdot \frac{\partial f}{\partial \mathbf{J}} \right].
 \end{align}
(c.f. equation (33) of \citealt{Kaufman1971}).


\subsection{The QL collision operator}
\label{sec:cql}

Our remaining task is to get explicit expressions for the vector $\mathbf{F}_g$ and the matrix $\mathsf{D}_g$ that appear in the QL collision operator $C_\mathrm{QL}$ (equation \eqref{eqn:CQL_drag_diffusion}). First, since we identified the first term in \eqref{eqn:dNgdt_twoparts} as $2\gamma_gN_g$ we must have $2\gamma_g = (2\pi)^3m^{-1}\int \md \bJ \, \mathbf{F}_g \cdot \partial f/\partial \bJ$.  We can compare this to \eqref{eqn:Landau} and consequently read off a consistent expression for\footnote{This is not the only consistent solution for $f$ but it is the only sensible one given that $f(\bJ)$ must vanish on the boundaries of action space and the result must hold for arbitrary $m$.}
$\mathbf{F}_g$:
\begin{align}
\label{eqn:F_g_explicit}
\mathbf{F}_g(\bJ) = - \frac{m}{\Lambda_g} \sum_{\bk} \bk \,  \delta (\bk\cdot \bm{\Omega} - \Omega_g)  \vert \phi_{g\bk}(\mathbf{J}) \vert^2. 
\end{align}
Second, we can compute $\mathsf{D}_g \equiv \tau^{-1}\langle \Delta \bJ \Delta \bJ \rangle_\tau^g$ by considering an encounter between a star, with initial coordinates $(\bm{\theta}_0, \bJ)$, and a wave quantum with frequency $\Omega_g$ and random phase $\alpha_g$. We calculate the first order change in the star's action $\delta \bJ(\tau)$, average $\delta \bJ \delta \bJ$ over all possible phases $\bm{\theta}_0$, $\alpha_g$ (c.f. equation (7) of \citealt{Hamilton2020}), divide by $\tau$, and take the limit $\tau \to \infty$. This calculation was already done for example in \S 3 of \citet{Binney1988} and in \S V of \citet{Kaufman1972}, so we relegate the derivation to Appendix \ref{sec:DQL_derivation} and only present the final result:
\begin{align}
\label{eqn:Dg}
  \mathsf{D}_g(\bJ)  &=  - \frac{\mathcal{E}_g}{\Omega_g \Lambda_gN_g}  \sum_\bk \bk\,\bk\, \delta (\bk\cdot \bm{\Omega} - \Omega_g) \vert \phi_{g\bk}(\mathbf{J}) \vert^2.
\end{align}

We may now plug \eqref{eqn:Dg} and \eqref{eqn:F_g_explicit} into \eqref{eqn:CQL_drag_diffusion} to get our final expression for the quasilinear collision operator:
\begin{align}
C_{\mathrm{QL}}[f,\{\mathcal{E}_g\}] &= - \frac{\partial }{\partial \bJ} \cdot 
 \sum_{\mathbf{k}} \mathbf{k}
 \sum_g  \delta(\bk\cdot\bm{\Omega}-\Omega_g) \frac{\vert\phi_{g\bk}(\mathbf{J}) \vert^2}{\Lambda_g}    \nn \\ &\times
  \left[m f(\mathbf{J}) + \frac{\mathcal{E}_g(t)}{\Omega_g} \, \bk \cdot \frac{\partial f}{\partial \mathbf{J}} \right].
  \label{eqn:C_QL}
\end{align}
(c.f. equation (39) of \citealt{Kaufman1971}). As expected, QL evolution is driven by resonances between a wave's oscillation frequency $\Omega_g$ and a star's dynamical frequency $\bk \cdot \bOm$.


\section{Discussion}
\label{sec:Discussion}

The unified kinetic theory consists of the kinetic equation for $f(\bJ)$:
\begin{equation}
\label{eqn:dfdt_final}
\frac{\partial f}{\partial t}
=
   C_{\mathrm{BL}}[f] + C_{\mathrm{QL}}[f,\{\mathcal{E}_g\}],
\end{equation}
with $C_\mathrm{BL}$ given in \eqref{eqn:BL} and $C_\mathrm{QL}$ given in \eqref{eqn:C_QL}, alongside the wave-kinetic equation for $\mathcal{E}_g$, equation \eqref{eqn:wave_kinetic}. The system is closed by specifying the initial value of $f(\bJ)$ and the initial wave energies $\mathcal{E}_g(t=0)$. We note that since the eigenmodes $\phi_g$ only appear in \eqref{eqn:wave_kinetic}, \eqref{eqn:C_QL} in the form $\vert \phi_{g\bk} \vert^2/\Lambda_g$, none of our results depend on their normalisation (see equation \eqref{eqn:Lambda_g}).


\subsection{Energy conservation}
\label{sec:energy}
We can now confirm that our closed system of equations conserves the total energy in mean field motion and waves:
\begin{align}
\label{eqn:Etot}
\mathcal{E}_\mathrm{tot}(t) \equiv (2\pi)^3 \int\md\mathbf{J} H(\mathbf{J})f(\mathbf{J},t) + \sum_g \mathcal{E}_g(t).
\end{align}
The rate of change of this energy is
\begin{align}
\label{eqn:roc_1}
  \frac{\md\mathcal{E}_\mathrm{tot} }{\md t} &= (2\pi)^3 \int\md\mathbf{J} H(\mathbf{J})\,C_{\mathrm{BL}}
  + (2\pi)^3 \int\md\mathbf{J} H(\mathbf{J})\,C_{\mathrm{QL}} \nn \\ &+  \sum_g \frac{\partial \mathcal{E}_g}{\partial t}.
\end{align}
It is well known from BL theory that the first term on the
right hand side is zero \citep{Heyvaerts2010,Chavanis2012b}.  Let us
define
\begin{equation}
\label{eqn:beta}
  \beta_{\bk}^g(\mathbf{J},t) \equiv - \frac{\vert \phi_{g\bk}(\mathbf{J}) \vert^2}{\Lambda_g}
  \left[mf(\mathbf{J}) + \frac{\mathcal{E}_g(t)}{\Omega_g} \, \bk \cdot \frac{\partial f}{\partial \mathbf{J}} \right],
\end{equation}
a factor common to both $C_\mathrm{QL}$ and $\partial \mathcal{E}_g/\partial t$. Then what remains of \eqref{eqn:roc_1} is the QL contribution:
\begin{align}
  \frac{\md\mathcal{E}_\mathrm{tot} }{\md t} =& (2\pi)^3\sum_{\bk}\sum_g \int\md\mathbf{J}
  \left[  H(\mathbf{J}) \bk \cdot \frac{\partial}{\partial \mathbf{J}} + \Omega_g \right] \nn \\ &\times
  \delta (\bk\cdot \bm{\Omega} - \Omega_g)  \beta^g_{\bk}(\mathbf{J},t).
\end{align}
Integrating the term involving $H(\bJ)$ by parts and discarding boundary terms, the square bracket can
be replaced by $[-\bk\cdot \bm{\Omega} + \Omega_g]$, but this is forced
to vanish by the presence of the delta function.  Thus, energy is conserved.  

Note that the friction and diffusion parts of the QL theory separately conserve energy.  For instance, if we ignored the friction part of $C_\mathrm{QL}$ and the spontaneous emission $S_g$ (as is often done in plasma kinetics for systems involving weakly unstable waves), then we would simply lose the term $mf(\bJ)$ from the right hand side of \eqref{eqn:beta}, but the energy conservation argument would work just the same.


\subsection{Recovering the QL and BL theories}

The right hand side of the kinetic equation \eqref{eqn:dfdt_final} is a superposition of the BL and QL collision operators $C_\mathrm{BL}$, $C_\mathrm{QL}$. 
In the limit where wave-star interactions dominate over the star-star interactions, one can set $C_\mathrm{BL}=0$ and recover the purely QL theory. Though we have insisted throughout that our system is stable $(\gamma_g < 0)$, our derivation also applies to weakly unstable systems $(\gamma_g > 0)$ as long as  $\gamma_g^{-1} \gg  \Omega_g \sim t_\mathrm{cross}$ \citep{Rogister1968}. The QL limit is often appropriate for weakly unstable systems like galactic disks wherein the perturbation due to growing waves quickly dwarfs the dressed noise \citep{Griv2002}. 

When considering hot systems like globular clusters one typically takes the opposite (BL) limit.  In this context it is argued that all waves decay on a timescale $\vert \gamma_g \vert^{-1}$ which is typically assumed to be $\sim t_\mathrm{cross}$, and that after this short time there are effectively no more wave-star interactions (e.g. Appendix F of \citealt{Fouvry2018}). However this is not quite correct, for two reasons.  First, we know that $\vert \gamma_g \vert^{-1}$ can be significantly longer than $t_\mathrm{cross}$ and that in intermediate-$N$ systems it might even become comparable to $t_\mathrm{relax}$ (\S\ref{sec:introduction}).  Second, it does not account for the fact that waves are \textit{always} present because they are continuously being spontaneously emitted by the system's discreteness noise (\S\ref{sec:waves}).  Indeed, we see from the wave kinetic equation \eqref{eqn:wave_kinetic} that in the absence of secular evolution, after a few damping times the wave energy $\mathcal{E}_g$ will have almost reached a steady state value:
\begin{align}
\label{eqn:steady_Eg}
    \mathcal{E}^\mathrm{steady}_g &= 
    \frac{\vert S_g\vert}{2\vert \gamma_g\vert } 
    \nn \\ 
    &= m\Omega_g \frac{\sum_{\bk} \int \md \bJ \,  \delta (\bk\cdot \bm{\Omega} - \Omega_g) \vert \phi_{g\bk}(\mathbf{J}) \vert^2 
 f(\bJ)
 }{\big \vert \sum_{\bk} \int \md \bJ \,  \delta (\bk\cdot \bm{\Omega} - \Omega_g) \vert \phi_{g\bk}(\mathbf{J}) \vert^2  \bk \cdot 
  \partial f/\partial \bJ \big\vert}.
\end{align}
Hence, the assumption we must actually make to recover the BL theory is that all $\gamma_g$ are sufficiently large that waves with energy $\mathcal{E}_g^\mathrm{steady}$ have a negligible effect on the diffusion of stars compared to that induced by steady-state dressed noise. For a fixed total mass of the system $M=Nm$ we have $\mathcal{E}^\mathrm{steady}_g \propto 1/N$, so equivalently this is a requirement that $N$ is sufficiently large that discreteness-driven waves are unimportant. 



\section{Conclusion}
\label{sec:conclusion}

In this paper we have derived a unified kinetic theory for stellar systems that accounts for the effects of both amplified noise and transient wave modes on secular evolution.  In the approximation we made (accounting for weakly damped waves only, ignoring mode coupling, and making the random phase approximation) the unified collision operator is essentially a superposition of the BL collision operator (describing dressed star-star interactions) and the QL collision operator (describing wave-star interactions).  The kinetic equation for the stars' phase space distribution $f$ must be complemented by a wave-kinetic equation which describes the energy $\mathcal{E}^g$ in each wave $g$. 

We call it `a' unified kinetic theory, rather than `the' unified kinetic theory, because throughout this paper we have made the assumption that waves are \textit{weakly damped} \eqref{eqn:weak_def}.  In practice this may be a poor approximation. For instance \citet{Weinberg1994,Murali1999} found that dipolar $\ell=1$ `sloshing' modes in spherical systems like globular clusters actually have $\gamma_\mathrm{dip.} \sim \Omega_\mathrm{dip.} \gg t_\mathrm{cross}$;  thus they are certainly \textit{long-lived} compared to the typical orbital period in the system, but not necessarily weakly damped in the sense of \eqref{eqn:weak_def}. An even more general theory might be required to incorporate these wave solutions.


\section*{Acknowledgements}
We thank A. Ellis for several helpful comments on the manuscript. CH is funded by a Science and Technology Facilities Council (STFC) Studentship.

\section*{Data availability}
No new data were generated or analysed in support of this research.




\bibliographystyle{mnras}
\bibliography{Bibliography} 

\appendix 

\section{Derivation of the QL diffusion tensor}
\label{sec:DQL_derivation}

Here we will derive the QL diffusion tensor $\mathsf{D}_g$ which we quoted in equation \eqref{eqn:Dg}. To do so we consider the interaction of a star with a bath of $N_g$ random wave quanta with frequency $\omega_g$. From the discussion in \S\ref{sec:wavestar} a generic wave quantum $j$ has potential
\begin{align}
\label{eqn:wave_potential}
\delta \Phi^j_g(\mathbf{x},t)  &= \frac{ A_g}{\sqrt{N_g}}\mathrm{Re}\left[\phi_g(\mathbf{x}) e^{-i\alpha^j_g} e^{-i\Omega_gt}\right] \nn
\\
&= \sum_\bk \delta \Phi^j_{g\bk}(\mathbf{J},t)e^{i\bk\cdot\bm{\theta}},
\end{align}
where the Fourier components are
\begin{align}
    \delta \Phi^j_{g\bk}(\mathbf{J},t) &\equiv \frac{A_g}{2\sqrt{N_g}} \nn \\ &\times \Big[ \phi_{g\bk}(\mathbf{J})e^{-i(\alpha_g^j + \Omega_g t)}
    +
    \phi_{g,-\bk}^*(\mathbf{J})e^{i(\alpha_g^j + \Omega_g t)}
    \Big].
\end{align}
Note that $\delta \Phi^j_{g\bk} = [\delta \Phi^j_{g,-\bk}]^*$, as must be the case since $\delta \Phi^j_{g}$ is real.  On the contrary $\phi_{g\bk} \neq \phi^*_{g,-\bk}$ in general since the eigenfunctions $\phi_g$ are complex. 

By analogy with \S\ref{sec:starstar} we now consider a test star with Hamiltonian $h=H(\bJ) + \delta \Phi^j_g$. From Hamilton's equation the change in the test star's action after time $\tau$ is:
\begin{align}
\delta \bJ (\bm{\theta}_0, \bJ, \Omega_g, \alpha_g^j, \tau) &= - \int_{0}^\tau \md t \, \frac{\p h}{\p \bm{\theta}} \nn \\ &= -  \mi \sum_{\bk} \bk \, \int_0^\tau \md t \, \delta \Phi^j_{g \bk}(\bJ,t) e^{i\bk\cdot (\bm{\theta}_0 + \bOm t)},
 \label{eqn:delta_J_first} 
\end{align}
where to get the second equality we substituted the mean field trajectory $\bJ(t) = \bJ(0)$,  $\bm{\theta}(t) = \bm{\theta}(0) + \bOm t \equiv \bm{\theta}_0 + \bOm t$. 

Next we compute the ensemble average $\langle \Delta \bJ \Delta \bJ\rangle_\tau^g$, which is the value of  $\Delta \bJ \Delta \bJ$ arising from the interaction of a star with a wave quantum after time $\tau$ (where $t_\mathrm{cross} \sim \Omega_g \ll \tau \ll \vert \gamma_g \vert^{-1}$) averaged over all wave quanta $j$.  To do so we replace the sum over $j$ with an integral, $N_g^{-1}\sum_{j=1}^{N_g} \to \int_0^{2\pi} \md \alpha_g/2\pi$, and integrate $\delta \bJ \delta \bJ$ over all possible initial angles $\bm{\theta}_0$ and phases $\alpha_g$ (c.f. equation (7) of \citealt{Hamilton2020}):
\begin{align}
   \langle \Delta \bJ \Delta \bJ \rangle^g_\tau 
    =& 
    \int \frac{\md \bm{\theta}_0}{(2\pi)^3} \frac{\md \alpha_g}{2\pi} 
    \nn
    \\ &\times \delta \bJ(\bm{\theta}_0, \bJ, \Omega_g, \alpha_g, \tau) \, \delta \bJ(\bm{\theta}_0, \bJ, \Omega_g, \alpha_g, \tau)
    \nn
    \\
    =& \sum_{\bk}\bk \,\bk \int_0^\tau \md t \int_0^\tau \md t'  \, e^{i\bk\cdot\bOm(t-t')}  c^g_\bk(\bJ,t-t'),
    \label{eqn:DeltaJDeltaJ_Binney}
\end{align}
where
\begin{align}
    c^g_\bk(\bJ,T) &\equiv  \frac{A_g^2}{4N_g}
    \Big[ 
    \vert \phi_{g\bk}(\mathbf{J}) \vert^2 e^{-i\Omega_g T}
    +
    \vert \phi_{g,-\bk}(\mathbf{J}) \vert^2 e^{i\Omega_gT}
    \Big].
\end{align}
Dividing \eqref{eqn:DeltaJDeltaJ_Binney} by $\tau$ and taking $\tau \to \infty$, we get (c.f. equations (3.8)-(3.9) of \citealt{Binney1988}):
\begin{align}
    \mathsf{D}_g = \frac{\pi A_g^2 }{2 N_g} \sum_\bk \bk \, \bk \, \delta(\bk\cdot \bOm - \Omega_g) \vert \phi_{g\bk}\vert^2,
\end{align}
which together with equation \eqref{eqn:Eg} gives the final expression \eqref{eqn:Dg}.


\label{lastpage}

\end{document}